\documentclass[aps,preprint,showpacs,floatfix]{revtex4}
\usepackage{bm,epsfig}
\begin{document}

\count255=\time\divide\count255 by 60 \xdef\hourmin{\number\count255}
  \multiply\count255 by-60\advance\count255 by\time
 \xdef\hourmin{\hourmin:\ifnum\count255<10 0\fi\the\count255}

\newcommand{\xbf}[1]{\mbox{\boldmath $ #1 $}}

\newcommand{\sixj}[6]{\mbox{$\left\{ \begin{array}{ccc} {#1} & {#2} &
{#3} \\ {#4} & {#5} & {#6} \end{array} \right\}$}}

\newcommand{\threej}[6]{\mbox{$\left( \begin{array}{ccc} {#1} & {#2} &
{#3} \\ {#4} & {#5} & {#6} \end{array} \right)$}}

\newcommand{\clebsch}[6]{\mbox{$\left( \begin{array}{cc|c} {#1} & {#2} &
{#3} \\ {#4} & {#5} & {#6} \end{array} \right)$}}

\newcommand{\iso}[6]{\mbox{$\left( \begin{array}{cc||c} {#1} & {#2} &
{#3} \\ {#4} & {#5} & {#6} \end{array} \right)$}}

\title{$1/N_c$ Corrections in Meson-Baryon Scattering}

\author{Herry J. Kwee}
\email{Herry.Kwee@asu.edu}

\author{Richard F. Lebed}
\email{Richard.Lebed@asu.edu}

\affiliation{Department of Physics and Astronomy, Arizona State
University, Tempe, AZ 85287-1504}

\date{August 2007}

\begin{abstract}
Corrections to meson/ground-state baryon scattering amplitudes in the
$1/N_c$ expansion of QCD have previously been shown to be controlled
by the $t$-channel difference $|I_t \! - \! J_t|$ of isospin and
angular momentum and by the change of hypercharge $Y_t$.  Here we
derive the corresponding expressions in the original scattering $s$
channel, allowing for nonzero meson spin and nontrivial SU(3) flavor
quantum numbers, and provide explicit examples of the crossing
relevant for $\pi N \! \to \rho N$ and $K N$ scattering.
\end{abstract}


\maketitle

\section{Introduction} \label{intro}

The $1/N_c$ expansion of large $N_c$ QCD, where $N_c$ is the number of
color charges, remains one of the very few predictive
model-independent schemes for studying strong interaction physics.  It
is particularly useful for studies of baryons, which require a content
of $N_c$ valence quarks in order to form color singlets.  Irreducible
operators containing many quark lines tend to be suppressed through
powers of $\alpha_s \! = \! O(1/N_c)$, meaning therefore that the
$1/N_c$ expansion produces an effective field theory for baryonic
systems.  This simple fact has been exploited to great effect for many
years.  The original examples using this ``operator''
technique~\cite{Prague} considered static baryon properties for
baryons stable under strong decays (or, more accurately, baryons whose
widths vanish as a power of $1/N_c$).  However, there is no reason to
expect baryonic dynamical properties, such as those probed in
meson-baryon scattering, to follow immediately from a static operator
approach.

Substantial progress derives from employing group theoretical
structures inspired by the Skyrme and other chiral soliton models,
since these models provide a natural way to couple baryons to chiral
mesons and to represent scattering processes.  Inspired by the famous
Adkins-Nappi-Witten papers~\cite{ANW}, the Siegen group~\cite{HEHW}
and Mattis and collaborators~\cite{Mattis,MM} exploited this group
theory to great phenomenological and formal effect in the 1980s.  The
key quantity in these analyses is a conserved ``grand spin'' ${\bf K}
\! = \! {\bf I} \! + \! {\bf J}$, which in soliton models
characterizes hedgehog states.  However, since $I$ and $J$ but not $K$
are externally measured quantum numbers, the physical baryon state
consists of a linear combination of $K$ eigenstates; in this approach
$K$ is therefore a good but hidden quantum number.

Nevertheless, the direct connection of these calculations to the large
$N_c$ limit, based partly upon the observation that baryons at large
$N_c$ are heavy and therefore semiclassical objects with the right
properties to be represented as solitons, remained indirect.  One
indication of this connection arises from the completely spin-flavor
symmetric quantum numbers of the lowest-lying baryons such as $N$ and
$\Delta$ (the ``ground-state band''), which is what one would expect
from states constructed entirely from a number ($N_c$) of $K \! = \!
0$ hedgehogic quarks.

Full compatibility with the large $N_c$ limit at the hadronic level
was shown in Ref.~\cite{CL1} to arise from ingredients already present
in the literature but not previously assembled.  First, in the
mid-1980s Donohue noted~\cite{Donohue} a great simplification of the
structure of meson-baryon scattering amplitudes when considered in the
$t$ rather than $s$ channel: The leading-order in $1/N_c$ amplitudes
for meson-baryon scattering [which are $O(N_c^0)$] have equal
$t$-channel isospin and angular momentum exchange quantum numbers,
$I_t \! = \! J_t$.  Mattis and Mukerjee then showed~\cite{MM} that a
group-theoretical crossing of the $I_t \! = \! J_t$ rule from the $t$
into the $s$ channel is equivalent to the ansatz of underlying $K$
conservation.  Years later, and in a different context, Kaplan and
Savage, and Kaplan and Manohar (KSM)~\cite{KapSavMan} showed that the
Dashen-Jenkins-Manohar consistency conditions~\cite{DJM}, which amount
to imposing unitarity order by order in powers of $1/N_c$ in
meson-baryon scattering, lead to the $I_t \! = \! J_t$ rule.  Finally,
just a few years ago, Ref.~\cite{CL1} assembled these facts to show
that the meson-baryon scattering results based upon the group theory
of chiral soliton models are in fact true results of the large $N_c$
limit.  Subsequent work showed~\cite{ItJt} that not only does the $I_t
\! = \!  J_t$ rule also apply to 3-flavor processes, but also
$Y_t \! = \! 0$ at leading $N_c$ order, due to interesting properties
of SU(3) Clebsch-Gordan coefficients (CGC) as $N_c$ grows large.

A recent and extensive body of
literature~\cite{CL1,ItJt,CDLN2,CDLM,CLchiral,CLpentSU3,CLSU3,CLSU3phenom,KL1,CLrest}
builds upon not only this connection to large $N_c$, but also upon the
observation that having an underlying conserved quantum number
contributing to multiple partial waves means that a resonant pole of a
given mass and width appearing in one partial wave of given $I$, $J^P$
quantum numbers also appears in numerous others, giving rise to
multiplets of baryon resonances degenerate in mass and width.  Here,
however, we are more interested in the structure of the amplitudes
themselves in terms of the $1/N_c$ expansion, rather than poles that
lie within them.

In particular, Ref.~\cite{CDLN2} pointed out that the KSM approach
also implies that amplitudes with $|J_t \! - \!  I_t| \! = \! n$ are
suppressed by at least $1/N_c^n$ compared to leading order, which
provides a way to study $1/N_c$-suppressed amplitudes---albeit
expressed in terms of $t$-channel quantities.  This classification was
employed phenomenologically to study $\pi N \! \to \pi N, \, \pi
\Delta$~\cite{CDLN2}, pion photoproduction~\cite{CDLM}, and chiral
threshold effects~\cite{CLchiral}, but always for spinless, nonstrange
mesons.  Generalizing this approach to mesons with spin (such as
$\rho$) and mesons with strangeness (such as $K$) are among the goals
of this work.

Moreover, in all previous cases the subleading amplitudes were
expressed in terms of $t$-channel quantum numbers, even though the $s$
channel is the most natural from the point of view of representing the
data (e.g., baryon resonances are identified in this channel).  It is
clearly highly desirable to express all amplitudes, including these
subleading effects, directly in terms of $s$-channel quantum numbers.
Deriving the exact nature of this transformation constitutes the
central goal of this paper.

The path to this goal may seem highly technical and mathematical, but
it has a simple physical significance.  Meson-baryon scattering can be
described in terms of a $1/N_c$ expansion, but previously the only
convenient way for distinguishing the various orders of $1/N_c$
corrections to the leading-order [$O(N_c^0)$] result was by describing
the process in terms of $t$-channel quantities.  Here we carry out the
crossing of the amplitudes into $s$-channel quantities in such a
manner that identifies quantum numbers whose changes correlate with
specific orders in the $1/N_c$ expansion, and as a result never lose
sight of where the various $t$-channel quantities contribute.

This paper is organized as follows: In Sec.~\ref{setup} we define the
relevant quantities of the scattering process.  Section~\ref{amprelns}
presents the original scattering amplitude expressions in the large
$N_c$ limit in terms of both $s$-channel and $t$-channel quantum
numbers.  In Sec.~\ref{crosssec} we obtain the relations for crossing
between $s$-channel and $t$-channel scattering descriptions,
independent of the $1/N_c$ expansion.  Section~\ref{1Ncorr} explains
how to incorporate $1/N_c$ corrections to meson-baryon scattering
processes in terms of $t$-channel quantum numbers.  Section~\ref{Main}
merges these ideas and shows how $1/N_c$ corrections may be expressed
in the original $s$-channel language, providing special cases and
examples.  Finally, Sec.~\ref{concl} summarizes.

\section{Observables}
\label{setup}

We consider meson-baryon scattering processes denoted by
\begin{equation} \label{sprocess}
\phi + B \to \phi^\prime + B^\prime .
\end{equation}
Here, $\phi \, (\phi^\prime)$ is a meson of spin $S_\phi \,
(S_{\phi^\prime})$ in the state of the flavor SU(3) representation
$R_\phi \, (R_{\phi^\prime})$ with isospin $I_\phi \,
(I_{\phi^\prime})$ and hypercharge $Y_\phi \, (Y_{\phi^\prime})$.  $B
\, (B^\prime)$ is a baryon of spin $S_B \, (S_{B^\prime})$
in the state of the flavor SU(3) representation $R_B \,
(R_{B^\prime})$ with isospin $I_B \, (I_{B^\prime})$ and hypercharge
$Y_B \, (Y_{B^\prime})$ within the ground-state spin-flavor multiplet
[the completely spin-flavor symmetric large $N_c$ generalization of
the SU(6) {\bf 56}, for which the nonstrange members $N$, $\Delta$,
etc., have $I_B \! = \! S_B$ and $Y_B \! = \! \frac{N_c}{3}$].  The
hadrons possess relative orbital angular momentum $L \, (L^\prime)$,
and their total spin angular momentum (not including relative orbital
angular momentum) is denoted $S \, (S^\prime)$.  Let us additionally
label the meson total angular momentum $J_\phi \, (J_{\phi^\prime})$,
and define the grand spin $K$ as the vector sum of $I_\phi$ and
$J_\phi$, and similarly for $K^\prime$.  Auxiliary quantum numbers
$\tilde K \, (\tilde K^\prime)$ label the vector sums of $L$ and
$I_\phi$ ($L^\prime$ and $I_{\phi^\prime}$).

We also label the intermediate compound $s$-channel state by total
quantum numbers $J_s$, $R_s$, $I_s$, and $Y_s$.  The representation
$R_s$ formed from $R_B \! \otimes \! R_\phi$ sometimes occurs more
than once in the product (for example, ${\bf 8} \! \otimes \! {\bf 8}$
contains two ${\bf 8}$'s), and this degeneracy quantum number---which
need not be the same in the initial and final state---is denoted by
$\gamma_s \, (\gamma^\prime_s)$.  Lastly, we define the compound
$t$-channel quantum numbers $I_t$ and $J_t$.  Classically, $I_t$ and
$J_t$ are vector differences of isospins and spins, respectively, of
the incoming and outgoing baryons; however, the simple difference
${\bf J}_1 \! - \! {\bf J}_2$ of two SU(2) vector operators does not
also transform as a vector operator.  Nevertheless, the proper
generalization is well known~\cite{Edmonds}: The notation $-{\bf J}$
is used as shorthand for the time-reversed form $\tilde{\bf J}$, an
operator whose eigenstates are related to those $( \left| J J_z
\right>)$ of ${\bf J}$ by $(-1)^{J+J_z} \left| J, - \! J_z \right>$.
Using this notation, the complete set of definitions reads
\begin{eqnarray}
& & {\bf I_s} \equiv {\bf I}_B \! + {\bf I}_\phi = {\bf I}_{B^\prime}
\! + {\bf I}_{\phi^\prime} \, , \nonumber \\ & &
{\bf S} \equiv {\bf S}_B \! + {\bf S}_\phi \, , \ \
{\bf S}^\prime \! \equiv {\bf S}_{B^\prime} \! + {\bf S}_{\phi^\prime}
\, , \nonumber \\ & &
{\bf J}_\phi \equiv {\bf L} \! + {\bf S}_\phi \, , \ \
{\bf J}_{\phi^\prime} \! \equiv {\bf L}^\prime \! +
{\bf S}_{\phi^\prime} \, , \nonumber \\ & &
{\bf J_s} \equiv {\bf L} + {\bf S} = {\bf J}_\phi \! + {\bf S}_B \! =
{\bf L}^\prime \! + {\bf S}^\prime = {\bf J}_{\phi^\prime} \! +
{\bf S}_{B^\prime}  , \nonumber \\ & &
\tilde {\bf K} \equiv {\bf L} \! + {\bf I}_\phi \, , \ \
\tilde {\bf K}^\prime \equiv {\bf L}^\prime \! + {\bf I}_{\phi^\prime}
, \nonumber \\ & &
{\bf K} \equiv \tilde {\bf K} + {\bf S}_\phi = {\bf I}_\phi + {\bf
J}_\phi = \tilde {\bf K}^\prime \! +
{\bf S}_{\phi^\prime} \! = {\bf I}_{\phi^\prime} \! +
{\bf J}_{\phi^\prime} \, , \nonumber \\ & &
{\bf I_t} \! \equiv \! - {\bf I}_B \! + {\bf I}_{B^\prime} =
{\bf I}_\phi - {\bf I}_{\phi^\prime} \, , \nonumber \\ & &
{\bf J_t} \equiv - {\bf S}_B \! + {\bf S}_{B^\prime} \! = {\bf J}_\phi
\! - {\bf J}_{\phi^\prime} \ ,
\label{vecdefs}
\end{eqnarray}
while hypercharges are additive,
\begin{eqnarray}
& & Y_s \equiv Y_B + Y_\phi =  Y_{B^\prime} + Y_{\phi^\prime}
\nonumber \\
& & Y_t \equiv -Y_B + Y_{B^\prime} = Y_\phi - Y_{\phi^\prime} \ ,
\end{eqnarray}
and $(R_s, \gamma_s) \! \in \! R_B \otimes R_\phi$, $(R_s,
\gamma^\prime_s) \! \in \! R_{B^\prime} \! \otimes R_{\phi^\prime}$.
Strictly speaking, the equality of initial- and final-state operators
(such as for ${\bf I_s}$) indicates the presence of conservation laws;
one may define, for example, a distinct ${\bf I_s^\prime}$ operator,
but barring explicit isospin violation in the scattering process, any
calculation shows the two operators to have the same effect.
Moreover, all of the definitions sum two SU(2) vector operators that
commute, a condition necessary in order for the resulting sum to obey
canonical commutation relations among its components.  The order of
summands for each definition has been carefully chosen to reflect the
order in which states are to be coupled: The couplings of $|J_1
J_{1z}\rangle |J_2 J_{2z} \rangle$ and $|J_2 J_{2z} \rangle |J_1
J_{1z} \rangle$ into a state $|J J_z \rangle$ differ by the phase
$(-1)^{J_1+J_2-J}$, while more nontrivial phases arise in
SU(3)~\cite{deSwart}.

The definitions of Eq.~(\ref{vecdefs}) must be considered carefully
from a physical point of view, because going beyond the large $N_c$
limit introduces recoil effects for the baryons.  Only when the
baryons are considered very heavy, in which case the center-of-mass
and rest frames of $B$ and $B^\prime$ coincide, does ${\bf J_t} \!
\equiv -{\bf S_B} \! + {\bf S_{B^\prime}}$ indicate the full change of
angular momentum of the baryon in the $t$ channel.  A full
relativistic treatment, as would be relevant to the most general case,
employs a helicity formalism; nevertheless, the quantum numbers
defined in Eq.~(\ref{vecdefs}) (particularly $J_t$) continue to be
well defined, even if their physical interpretation is not so simple
as in the heavy-baryon limit.  Indeed, this is one motivation for
re-expressing the $t$-channel amplitudes in terms of the more familiar
$s$-channel quantum numbers.

\section{Scattering Amplitude Relations} \label{amprelns}

The original derivations~\cite{HEHW,Mattis,MM} of linear relations
among meson-baryon scattering amplitudes rely upon solitonic
representations of the baryon wave function.  In particular, the
underlying state is given by a hedgehog configuration, whose
functional dependence upon coordinates appears only through the
characteristic mixed space-isospin inner product $\hat{\bf r} \cdot
{\xbf \tau}$; it is therefore an eigenstate of neither spin nor
isospin separately, but rather the vector sum ${\bf K} \! \equiv \!
{\bf I} \!  + \!  {\bf J}$ of the two.  Each value of $K$ gives rise
to a distinct soliton configuration, which may be probed through
scattering with mesons; a distinct {\it reduced\/} amplitude $\tau$
occurs for each $K$ and initial (final) value $L \, (L^\prime)$ of
relative orbital angular momentum.  If the meson probes also carry
spin, one must also include the auxiliary quantum numbers $\tilde{K}$,
$\tilde{K}^\prime$ defined in Eq.~(\ref{vecdefs}), and if they carry
strangeness, then one must also include isospin and hypercharge
quantum numbers.  The most general such reduced amplitude carries the
labels $\tau^{ \{ I I^\prime Y \} }_{K \tilde{K} \tilde{K}^\prime L
L^\prime}$.

From here it is a straightforward albeit tedious exercise to use the
definitions in Eq.~(\ref{vecdefs}) for coupling all appropriate
quantum numbers to represent a full physical $S$ matrix scattering
amplitude $S_{L L^\prime S S^\prime J_s R_s \gamma_s \gamma^\prime_s
I_s Y_s}$ (dependence upon particular $B, \, B^\prime, \, \phi, \,
\phi^\prime$ quantum numbers being implicit).  The physical baryon
state is given by the linear combination of solitonic configurations
such that the composite state is the appropriate eigenstate of isospin
and spin.  $S_{L L^\prime S S^\prime J_s R_s \gamma_s \gamma^\prime_s
I_s Y_s}$ is reduced (i.e., independent of $I_z, \, J_z$ quantum
numbers) in the sense of the Wigner-Eckart theorem.  The full
expression, first derived in Ref.~\cite{MM} and corrected in
Ref.~\cite{CLpentSU3}, reads
\begin{eqnarray}
\lefteqn{S_{L L^\prime S S^\prime J_s R_s \gamma^{\vphantom\prime}_s
\gamma^\prime_s I_s Y_s}} \nonumber \\
& = & (-1)^{S_B - S_{B^\prime}}
([R_B][R_{B^\prime}][S][S^\prime])^{1/2} / [R_s]
\sum_{\stackrel{\scriptstyle I \in R_\phi, \; I^\prime \in
R_{\phi^\prime},}{I^{\prime\prime} \in R_s, \; Y \in R_\phi \cap
R_{\phi^\prime}}} (-1)^{I + I^\prime + Y} [I^{\prime\prime}]
\nonumber \\ & & \times
\left( \begin{array}{cc||c} R_B & R_\phi & R_s \, \gamma_s \\ S_B
\frac{N_c}{3} & I Y & I^{\prime\prime} \, Y \! \! + \! \frac{N_c}{3}
\end{array} \right)
\left( \begin{array}{cc||c} R_B & R_\phi & R_s \, \gamma_s \\ I_B
Y_B & I_\phi Y_\phi & I_s Y_s \end{array} \right) \nonumber \\
& & \times
\left( \begin{array}{cc||c} R_{B^\prime} & R_{\phi^\prime} & R_s \,
\gamma^\prime_s \\ S_{B^\prime}
\frac{N_c}{3} & I^\prime Y & I^{\prime\prime} \, Y \! \! + \!
\frac{N_c}{3}
\end{array} \right)
\left( \begin{array}{cc||c} R_{B^\prime} & R_{\phi^\prime} & R_s \,
\gamma^\prime_s \\ I_{B^\prime} Y_{B^\prime} & I_{\phi^\prime}
Y_{\phi^\prime} & I_s Y_s \end{array} \right)
\nonumber \\ & & \times \sum_{K, \tilde{K} , \tilde{K}^\prime}
[K] ([\tilde{K}][\tilde{K}^\prime])^{1/2}
\left\{ \begin{array}{ccc}
L   & I                & \tilde{K} \\
S   & S_B              & S_\phi    \\
J_s & I^{\prime\prime} & K \end{array} \right\}
\! \left\{ \begin{array}{ccc}
L^\prime & I^\prime         & \tilde{K}^\prime \\
S^\prime & S_{B^\prime}     & S_{\phi^\prime} \\
J_s      & I^{\prime\prime} & K \end{array} \right\}
\tau^{\left\{ I I^\prime Y \right\}}_{K \tilde{K} \tilde{K}^\prime \!
L L^\prime} \ ,
\label{schannel3}
\end{eqnarray}
where the double-barred quantities are SU(3) isoscalar
factors~\cite{CLSU3}, quantities $[X]$ indicate representation
multiplicities (for example, for angular momenta $[J] \! = \! 2J \! +
\! 1$), and the braced quantities are standard SU(2) $9j$ symbols.

In comparison, the original 2-flavor result~\cite{Mattis} reads
\begin{eqnarray}
S_{L L^\prime S S^\prime I_s J_s} & = & \sum_{K, \tilde{K} ,
\tilde{K}^\prime} [K]
([S_B][S_{B^\prime}][S][S^\prime][\tilde{K}][\tilde{K}^\prime])^{1/2}
\nonumber \\
& & \times \left\{ \begin{array}{ccc}
L   & I_\phi & \tilde{K} \\
S   & S_B   & S_\phi    \\
J_s & I_s   & K \end{array} \right\}
\left\{ \begin{array}{ccc}
L^\prime & I_{\phi^\prime} & \tilde{K}^\prime \\
S^\prime & S_{B^\prime}    & S_{\phi^\prime} \\
J_s      & I_s             & K \end{array} \right\}
\tau_{K \tilde{K} \tilde{K}^\prime L L^\prime} . \label{schannel2}
\end{eqnarray}
where~\cite{CLSU3phenom} $\tau_{K \tilde{K} \tilde{K}^\prime L
L^\prime} \! \equiv \!  (-1)^{I_B - I_{B^\prime} + I_\phi -
I_{\phi^\prime}} \tau^{\{ I_\phi I_{\phi^\prime} Y_\phi \}}_{K
\tilde{K} \tilde{K}^\prime L L^\prime}$.  Also shown in
Ref.~\cite{CLSU3phenom} is the manner in which the flavor SU(3)
factors reduce to isospin SU(2) factors in the large $N_c$ limit.
Such factors are nontrivial because the SU(3) representations and
isoscalar factors are $N_c$ dependent; for example, the nucleon
resides not in a literal {\bf 8} = (1,1) in the usual weight notation,
but rather ``{\bf 8}'' = [$1, \, (N_c \! - \! 1)/2$].

One may also opt to express these amplitude expressions using
$t$-channel quantities.  Now the process is no longer expressed in the
quantum numbers relevant to the $s$-channel process
Eq.~(\ref{sprocess}), but those of the corresponding $t$-channel
process
\begin{equation} \label{tprocess}
\phi + \bar \phi^\prime \to \bar B + B^\prime \ .
\end{equation}
In particular, the quantum numbers $S, \, S^\prime, \, I_s, \, J_s$
are traded for $J_\phi, \, J_\phi^\prime, \, I_t, \, J_t$.  Of course,
the kinematic region for the literal on-shell process of
Eq.~(\ref{tprocess}) [which requires momentum transfers obtained from
the very large value $t \! \ge \! (m_{\bar B} \! + \! m_{B^\prime})^2
\! = \! O(N_c^2)$] is very different from the one of
Eq.~(\ref{sprocess}) [which only requires $O(N_c^0)$ momentum
transfers].  Moreover, standard $N_c$ power counting~\cite{Witten}
shows that, while meson-baryon scattering amplitudes are $O(N_c^0)$,
those for $\bar B B$ production are suppressed as $e^{-N_c}$.
Nevertheless, the usual field-theoretic assumptions hold that the same
amplitude (via analytic continuation of momenta) appears in both
regions.  We are interested only in the behavior of on-shell
$s$-channel processes expressed in terms of $t$-channel quantum
numbers.

This problem was originally addressed in Ref.~\cite{MM}; here we
present corrected versions~\cite{ItJt,phasecorr} of the expressions
derived in that work.  In the 3-flavor case, one finds
\begin{eqnarray}
\lefteqn{S_{L L^\prime J_{\phi^{\vphantom\prime}} J_{\phi^\prime} J_t
R_t \gamma^{\vphantom\prime}_t \gamma^\prime_t I_t Y_t} =
(-1)^{S_{\phi^\prime} - S_{\phi^{\vphantom\prime}} + 
J_{\phi^{\phantom\prime}} - J_t}
([R_{B^{\vphantom\prime}}][R_{B^\prime}][J_{\phi^{\vphantom\prime}}]
[J_{\phi^\prime}])^{1/2} / [R_t]}
\nonumber \\ & \times &
\sum_{\stackrel{\scriptstyle I \in R_{\phi^{\vphantom\prime}}, \,
I^\prime \in R_{\phi^\prime} \! ,}{Y \in R_{\phi^{\vphantom\prime}}
\cap \, R_{\phi^\prime}}} 
\left( \begin{array}{cc||c} R_{\phi^{\vphantom\prime}} &
R^*_{\phi^\prime} & R_t \, \gamma_t \\ I Y & I^\prime \! , \!
- \! Y & J_t \, 0 \end{array} \right)
\left( \begin{array}{cc||c} R_{\phi^{\vphantom\prime}} &
R^*_{\phi^\prime} & R_t \, \gamma_t \\ I_\phi Y_\phi
& I_{\phi^\prime}, \! - \! Y_{\phi^\prime} & I_t \, Y_t
\end{array} \right) \nonumber \\ & \times &
\left( \begin{array}{cc||c} R^*_B & R_{B^\prime} & R_t \,
\gamma^\prime_t \\ S_B, - \frac{N_c}{3} & S_{B^\prime} \! ,
+ \frac{N_c}{3} & J_t \, 0 \end{array} \right)
\left( \begin{array}{cc||c} R^*_B & R_{B^\prime} & R_t \,
\gamma^\prime_t \\ I_B, - \! Y_B & I_{B^\prime} Y_{B^\prime} & I_t \,
Y_t \end{array} \right)
\nonumber \\ & \times & \sum_{K, \tilde{K}, \tilde{K}^\prime}
(-1)^{K + \tilde{K}^\prime - \tilde{K} - \frac{Y}{2}} [K]
( [\tilde{K}] [\tilde{K}^\prime] )^{1/2}
\left\{ \begin{array}{ccc}
J_\phi   & I              & K \\
I^\prime & J_{\phi^\prime} & J_t \end{array} \right\} \!
\left\{ \begin{array}{ccc}
J_\phi    & I      & K \\
\tilde{K} & S_\phi & L \end{array} \right\} \!
\left\{ \begin{array}{ccc}
J_{\phi^\prime}  & I^\prime        & K \\
\tilde{K}^\prime & S_{\phi^\prime} & L^\prime \end{array} \right\} \!
\nonumber \\ & \times &
\tau^{\left\{ I I^\prime Y \right\}}_{K \tilde{K} \tilde{K}^\prime L
L^\prime} \ ,
\label{tchannel3}
\end{eqnarray}
which reduces in the 2-flavor case, thanks to the 3-flavor $I_t \! =
\! J_t$ and $Y_t \! = \! 0$ rules~\cite{ItJt}, to
\begin{eqnarray}
\lefteqn{S_{L L^\prime J^{\vphantom\prime}_\phi J_{\phi^\prime} I_t
J_t} = \delta_{I_t J_t} (-1)^{
+ I_{\phi^{\prime}} + S_{\phi^\prime} 
- S_{\phi^{\vphantom\prime}} + J_{\phi^{\vphantom\prime}}
- J_{t^{\vphantom\prime}}}
([S_{B^{\vphantom\prime}}][S_{B^\prime}] [J_{\phi^{\vphantom\prime}}]
[J_{\phi^\prime}])^{1/2} / [I_t]} \nonumber \\
& \times & \sum_{K, \tilde{K}, \tilde{K}^\prime}
(-1)^{K + \tilde{K}^\prime - \tilde{K}} [K]
( [\tilde{K}] [\tilde{K}^\prime] )^{1/2}
\left\{ \begin{array}{ccc}
J_\phi          & I_\phi          & K \\
I_{\phi^\prime} & J_{\phi^\prime} & J_t \end{array} \right\} \!
\left\{ \begin{array}{ccc}
J_\phi    & I_\phi & K \\
\tilde{K} & S_\phi & L \end{array} \right\} \!
\left\{ \begin{array}{ccc}
J_{\phi^\prime}  & I_{\phi^\prime} & K \\
\tilde{K}^\prime & S_{\phi^\prime} & L^\prime \end{array} \right\} \!
\nonumber \\ & \times &
\tau_{K \tilde{K} \tilde{K}^\prime L L^\prime} \ .
\label{tchannel2}
\end{eqnarray}

Built into each of these expressions is the solitonic baryon wave
function, which strictly speaking is adequate only in the large $N_c$
limit.  Each expression carries the correct $O(N_c^0)$ scaling as long
as the reduced amplitudes $\tau$ are also $O(N_c^0)$.  When $1/N_c$
corrections are included, the corrections are of two types: $O(1/N_c)$
corrections to the amplitudes $\tau$ [which are multiplicative in
nature and do not change the group-theoretical structures exhibited in
Eqs.~(\ref{schannel3})--(\ref{tchannel2}); for example, corrections to
the profile functions of soliton models are of this type] and $1/N_c$
corrections with group-theoretical structures different from those in
Eqs.~(\ref{schannel3})--(\ref{tchannel2}).  Since $I_t \! = \! J_t$
(and $Y_t \! = \! 0$) is a direct consequence of these calculations,
it follows that amplitudes with $I_t \! \neq \! J_t$ are necessarily
subleading in $1/N_c$~\cite{CDLN2}.

We emphasize that solitonic wave functions, although part of the
original derivations, are not essential to the process.  They are
invoked here merely to describe the historical path by which such
relations first appeared.  In the general large $N_c$ approach, the
only essential feature of the reduced amplitudes $\tau$'s is that they
are $O(N_c^0)$.

We close this section with a note on how the $I_t \! = \! J_t$ rule
arises when one derives Eq.~(\ref{tchannel2}) directly, rather than
through a reduction of the 3-flavor result.  There, the result holds
almost trivially: The solitonic baryon wave functions of good quantum
numbers $I_B \!  = \! S_B$ are obtained~\cite{ANW,HEHW,Mattis,MM} by
rotating the canonical soliton through an SU(2) element $A$ using the
rotation matrix $D^{(I_B=S_B)}_{I_{Bz}, \, -S_{Bz}} (A)$ of rank $I_B
\!  = \!  S_B$.  In the $t$ channel one has such a rotation matrix for
$\bar B$ and one for $B^\prime$.  When these are combined using the
standard identity~\cite{Edmonds}
\begin{eqnarray}
\lefteqn{D^{(I_{\bar B} = S_{\bar B})}_{-I_{Bz}, S_{Bz}} \! (A) \,
D^{(I_{B^\prime} = S_{B^\prime})}_{I_{B^\prime \! z},
-S_{B^\prime \! z}} (A)} & & \nonumber \\
& = & \sum_{\cal J} \clebsch{I_{\bar B}}{I_{B^\prime}}{\cal
J}{-I_{Bz}}{I_{B^\prime \! z}} {-I_{Bz}+I_{B^\prime \! z}}
\clebsch{S_{\bar B}}{S_{B^\prime}}{\cal J}{S_{Bz}}{-S_{B^\prime
z}}{S_{Bz} \! - S_{B^\prime \! z}} D^{({\cal J})}_{-I_{Bz} \! +
I_{B^\prime \! z}, \, S_{Bz} - S_{B^\prime \! z}} (A) \, ,
\end{eqnarray}
the same value of ${\cal J}$ occurs in both the isospin and spin CGC.
In light of the definitions of ${\bf I_t}$ and ${\bf J_t}$ in
Eq.~(\ref{vecdefs}), this expression shows that ${\cal J}
\! = \! I_t \! = \! J_t$, a result that follows directly from the
spin-flavor symmetry of the soliton.

\section{Crossing Relations} \label{crosssec}

Two equivalent approaches lead to the amplitudes of
Eq.~(\ref{tchannel3}) or (\ref{tchannel2}): One may work directly with
the process written in terms of $t$-channel states, as in
Eq.~(\ref{tprocess}), or derive a generic expression for crossing from
the $s$ to the $t$ channel.  The latter approach, pioneered in
Ref.~\cite{RS}, was also employed in Ref.~\cite{MM}.  Since this is
the means by which one may express the subleading in $1/N_c$
amplitudes in terms of $s$-channel quantities, we take some care to
explain its derivation relevant to the present case.

\subsection{2-Flavor Case}
As seen in the previous section, the spin-only angular momenta ${\bf
S}, \, {\bf S}^{\prime}$ are useful in the $s$ channel, while the
meson-only angular momenta ${\bf J}_{\phi^{\vphantom\prime}} , \, {\bf
J}_{\phi^\prime}$ are useful in the $t$ channel.  Indeed, transforming
between one order of coupling and another is precisely the original
purpose of $6j$ symbols~\cite{Edmonds}:
\begin{equation} \label{baserecouple}
\left| j_1, (j_2 j_3) \, J_{23}, JM \right> = \sum_{J_{12}} \left|
(j_1 j_2) \, J_{12}, j_3, JM \right>
(-1)^{j_1 + j_2 + j_3 + J} \sqrt{[J_{12}][J_{23}]}
\sixj{j_1}{J_{12}}{j_2}{j_3}{J_{23}}{J} \, .
\end{equation}
In our case, $j_1 \! \to \! S_B$, $j_2 \! \to \! S_\phi$, $j_3 \! \to
\! L$, $J_{12} \! \to \! S$, $J_{23} \! \to \! J_\phi$, and
$J \! \to \! J_s$, with analogous assignments for the primed
quantities.  Additional phases $(-1)^{L + S - J_s}$, $(-1)^{L + S_\phi
- J_\phi}$, and $(-1)^{J_\phi + S_B - J_s}$ arise from noting that the
orders of coupling ${\bf S} + {\bf L} \to {\bf J_s}$, ${\bf S_\phi} \!
+ {\bf L} \to {\bf J_\phi}$, and ${\bf S_B} \! + {\bf J_\phi} \to {\bf
J_s}$, respectively, as given in Eq.~(\ref{baserecouple}) are opposite
those given in Eq.~(\ref{vecdefs}).  Using the symmetry properties of
$6j$ symbols (invariance under exchanging two columns or the upper and
lower entries of any two rows), one finds
\begin{eqnarray}
\lefteqn{S_{L L^\prime J_{\phi^{\vphantom\prime}} J_{\phi^\prime}
I_s J_s} = \sum_{S,S^\prime}
\sqrt{[S][S^\prime][J_{\phi^{\vphantom\prime}}][J_{\phi^\prime}]}
} & & \nonumber \\ & \times &
(-1)^{L - L^\prime + S - S^\prime}
\sixj{J_s}{J_\phi}{S_B}{S_\phi}{S}{L}
\sixj{J_s}{J_{\phi^\prime}}{S_{B^\prime}}{S_{\phi^\prime}}{S^\prime}
{L^\prime} S_{L L^\prime S^{\vphantom\prime} S^\prime I_s J_s} \ .
\label{recouple}
\end{eqnarray}

The next step is to cross the $t$-channel quantities to an $s$-channel
description~\cite{RS}.  The crossing at the computational level, in
light of Eq.~(\ref{tprocess}), consists first of establishing a phase
convention for exchanging bras with kets, and second of moving the CGC
associated with $\bar \phi^\prime$ and $\bar B$---the two that form
$I_t$ and the two that form $J_t$ according to the last two
definitions of Eq.~(\ref{vecdefs})---to the $s$-channel side of the
equation.  The phase convention for two flavors is
\begin{equation}
\left| I I_z \right> \leftrightarrow (-1)^{I + I_z} \left< I - I_z
\right| \, , \label{reflect2}
\end{equation}
and for three flavors we choose
\begin{equation} \label{reflect3}
\left| R, I I_z, Y \right> \leftrightarrow (-1)^{I_z + \frac{Y}{2} -
\frac{1}{3}(2p+q)} \left< R^* \! , I -I_z, -Y \right| \, ,
\end{equation}
where the SU(3) representation $R$ in weight notation is $(p,q)$.  The
SU(2) CGC are moved from one side of an equation to the other by means
of their orthogonality relations, which leads to four CGC for isospin
and four for spin, summed over all magnetic quantum numbers.  But such
invariants are again $6j$ symbols; specifically, one finds
\begin{eqnarray}
\lefteqn{S_{L L^\prime J_{\phi^{\vphantom\prime}}
J_{\phi^\prime} I_t J_t} = \sum_{I_s, J_s} [I_s][J_s]
(-1)^{I_s + I_t + S_B - I_{\phi^\prime}}
(-1)^{J_s + J_t + S_B - J_{\phi^\prime}}}
& & \nonumber \\ & \times &
\sixj{S_{B^\prime}}{S_B}{I_t}{I_\phi}{I_{\phi^\prime}}{I_s}
\sixj{S_{B^\prime}}{S_B}{J_t}{J_\phi}{J_{\phi^\prime}}{J_s}
S_{L L^\prime J_{\phi^{\vphantom\prime}} J_{\phi^\prime} I_s J_s}
\, .
\label{crossonly}
\end{eqnarray}
Combining Eqs.~(\ref{recouple}) and (\ref{crossonly}) then gives
\begin{eqnarray}
\lefteqn{S_{L L^\prime J_{\phi^{\vphantom\prime}} J_{\phi^\prime}
I_t J_t} = \sum_{S,S^\prime,I_s,J_s} [I_s] [J_s]
([J_{\phi^{\vphantom\prime}}][J_{\phi^\prime}] [S][S^\prime])^{1/2}
(-1)^{I_t + I_s + J_t + J_s + L - L^\prime
+ S - S^\prime + 2S_B
- J_{\phi^{\prime}} - I_{\phi^\prime}}}
& & \nonumber \\ & \times &
\sixj{S_{B^\prime}}{S_B}{I_t}{I_\phi}{I_{\phi^\prime}}{I_s}
\sixj{S_{B^\prime}}{S_B}{J_t}{J_\phi}{J_{\phi^\prime}}{J_s}
\sixj{J_s}{J_\phi}{S_B}{S_\phi}{S}{L}
\sixj{J_s}{J_{\phi^\prime}}{S_{B^\prime}}{S_{\phi^\prime}}
{S^\prime}{L^\prime} S_{L L^\prime S S^\prime I_s J_s} \ .
\label{crossreln1}
\end{eqnarray}
Apart from a corrected phase, this expression was first derived in
Ref.~\cite{MM}.  Moreover, using the orthogonality properties of $6j$
symbols~\cite{Edmonds}, Eq.~(\ref{crossreln1}) may be inverted to give
an inverse expression with precisely the same $6j$ symbols and phases:
\begin{eqnarray}
\lefteqn{S_{L L^\prime S S^\prime I_s J_s} =
\sum_{J_{\phi^{\vphantom\prime}},J_{\phi^\prime},I_t,J_t} [I_t] [J_t]
([J_{\phi^{\vphantom\prime}}]
[J_{\phi^\prime}] [S][S^\prime])^{1/2}
(-1)^{I_t + I_s + J_t + J_s + L - L^\prime
+ S - S^\prime + 2S_B - J_{\phi^{\prime}} - I_{\phi^\prime}}}
& & \nonumber \\ & \times &
\sixj{S_{B^\prime}}{S_B}{I_t}{I_\phi}{I_{\phi^\prime}}{I_s}
\sixj{S_{B^\prime}}{S_B}{J_t}{J_\phi}{J_{\phi^\prime}}{J_s}
\sixj{J_s}{J_\phi}{S_B}{S_\phi}{S}{L}
\sixj{J_s}{J_{\phi^\prime}}{S_{B^\prime}}{S_{\phi^\prime}}
{S^\prime}{L^\prime} S_{L L^\prime J_{\phi^{\vphantom\prime}}
J_{\phi^\prime} I_t J_t} \ .
\label{crossreln2}
\end{eqnarray}

\subsection{3-Flavor Case}
Unfortunately, the 3-flavor crossing expressions for a process with
external particles of fixed $I$ and $Y$ is not so straightforward to
express in a simple closed form similar to Eq.~(\ref{crossreln2}).
The orthogonality relations for SU(2) CGC used to prove
Eq.~(\ref{crossonly}) sum over the SU(2) magnetic quantum numbers
$I_z$ and $J_z$ but not the Casimirs $I$ and $J$.  However, in the
case of flavor SU(3) the quantum numbers summed in the analogous CGC
orthogonality relations~\cite{deSwart} also include values of $I$ and
$Y$, while in a given physical process (e.g., $K N$ rather than $\pi
\Lambda$), these values are specified by particular external states and
are not summed.  The 3-flavor crossing conditions must be written as a
set of linear equations, with SU(3) isoscalar factors [those of
Eqs.~(\ref{schannel3}) and (\ref{tchannel3})] on the two sides.

This result means that one cannot present an explicit expression for
the crossing of a particular amplitude for which the SU(3) quantum
numbers $R_s \gamma_s$ are specified.  Fortunately, physical data
specify quantum numbers such as $I_s$, $J_s$, and $L$, but not whether
the scattering proceeds through an octet channel, for example. In
fact, the 2-flavor crossing relations remain useful, for one may
eliminate the SU(3) behavior simply by summing over all possible
intermediate SU(3) quantum numbers, weighted by the appropriate SU(3)
isoscalar factors~\cite{RS3flavor}.  Note that these are not just
trivial projections from strange to nonstrange amplitudes, but rather
weighted averages of strangeness-containing amplitudes for which the
SU(3) quantum numbers are irrelevant.  Let us define in this way pure
SU(2) amplitudes $\bar S_{L L^\prime S S^\prime I_s J_s}$ and $\bar
S_{L L^\prime J_{\phi^{\vphantom\prime}} J_{\phi^\prime} I_t J_t}$:
\begin{equation} \label{Sbars}
\bar S_{L L^\prime S S^\prime I_s J_s}
\equiv \sum_{R_s, \gamma^{\vphantom\prime}_s, \gamma^\prime_s}
\left( \begin{array}{cc||c} R_B & R_\phi & R_s \, \gamma_s \\ I_B
Y_B & I_\phi Y_\phi & I_s Y_s \end{array} \right)
\left( \begin{array}{cc||c} R_{B^\prime} & R_{\phi^\prime} & R_s \,
\gamma^\prime_s \\ I_{B^\prime} Y_{B^\prime} & I_{\phi^\prime}
Y_{\phi^\prime} & I_s Y_s \end{array} \right)
S_{L L^\prime S S^\prime J_s R_s \gamma^{\vphantom\prime}_s
\gamma^\prime_s I_s Y_s} \, ,
\end{equation}
\begin{equation} \label{Sbart}
\bar S_{L L^\prime J_{\phi^{\vphantom\prime}}
J_{\phi^\prime} I_t J_t} \equiv \sum_{R_t, \gamma^{\vphantom\prime}_t,
\gamma^\prime_t}
\left( \begin{array}{cc||c} R_{\phi^{\vphantom\prime}} &
R^*_{\phi^\prime} & R_t \, \gamma_t \\ I_\phi Y_\phi
& I_{\phi^\prime}, \! - \! Y_{\phi^\prime} & I_t \, Y_t
\end{array} \right)
\left( \begin{array}{cc||c} R^*_B & R_{B^\prime} & R_t \,
\gamma^\prime_t \\ I_B, - \! Y_B & I_{B^\prime} Y_{B^\prime} & I_t \,
Y_t \end{array} \right)
S_{L L^\prime J_{\phi^{\vphantom\prime}} J_{\phi^\prime} J_t
R_t \gamma^{\vphantom\prime}_t \gamma^\prime_t I_t Y_t} \, .
\end{equation}
Then the relation between the isospin amplitudes $\bar S$ is the same
as for the original SU(2) amplitudes $S$ in Eq.~(\ref{crossreln1}),
except with the SU(2) crossing phases in Eq.~(\ref{reflect2}) replaced
by the ones for SU(3) given in Eq.~(\ref{reflect3}):
\begin{eqnarray}
\lefteqn{\bar S_{L L^\prime J_{\phi^{\vphantom\prime}}
J_{\phi^\prime} I_t J_t} =
\sum_{S,S^\prime,I_s,J_s} [I_s] [J_s]
([J_{\phi^{\vphantom\prime}}][J_{\phi^\prime}] [S][S^\prime])^{1/2}}
& & \nonumber \\ & \times &
(-1)^{I_t + I_s + J_t + J_s
+ L - L^\prime + S - S^\prime + 2S_B - J_{\phi^\prime}
+ Y_{\phi^\prime}/2}
\nonumber \\ & \times &
\sixj{S_{B^\prime}}{S_B}{I_t}{I_\phi}{I_{\phi^\prime}}{I_s}
\sixj{S_{B^\prime}}{S_B}{J_t}{J_\phi}{J_{\phi^\prime}}{J_s}
\sixj{J_s}{J_\phi}{S_B}{S_\phi}{S}{L}
\sixj{J_s}{J_{\phi^\prime}}{S_{B^\prime}}{S_{\phi^\prime}}
{S^\prime}{L^\prime} \bar S_{L L^\prime S S^\prime I_s J_s} \ .
\label{crossreln3}
\end{eqnarray}
We see that a direct inversion of Eqs.~(\ref{Sbars}) and (\ref{Sbart})
is not possible, because the isoscalar factor orthogonality relations
required to do so sum over externally fixed quantum numbers such as
$I_B, \, Y_B$.  However, as noted previously the full 3-flavor
amplitudes themselves depend (implicitly) on these quantum numbers;
only if one assumes the amplitudes are the same for all states in the
SU(3) multiplets may one perform such an inversion and express the
3-flavor crossing relation in closed form.  If one is unwilling to
embrace this level of SU(3) symmetry but insists on retaining all
SU(3) quantum numbers, the best one can do for a given process is
obtain linear relations between the amplitudes expressed in the $s$
and $t$ channels.  Fortunately, as discussed above, this degree of
specificity is unnecessary for comparison with data; in
Sec.~\ref{Main} we see that only Eq.~(\ref{crossreln3}) is required to
study, for example, $KN$ scattering.

\section{The $I_t \! = \! J_t$ Rule and Its Corrections}
\label{1Ncorr}

Of course, in nature $N_c$ is only 3, and a robust phenomenological
analysis is not possible unless the structure of $1/N_c$ corrections
is understood.  As we have seen, the scattering amplitude expressions
based upon chiral soliton models are quite impressive, but
nevertheless hold only in the large $N_c$ limit.  To move beyond this
point one requires additional input, which is provided by the operator
approach.  Starting with the ansatz (common to both soliton and quark
models) that ground-state band baryons are completely symmetric under
the combined spin-flavor symmetry, one divides the baryon wave
function into $N_c$ quark interpolating fields, each of which carries
spin, flavor, and color fundamental representation indices.  The color
index, completely antisymmetrized among the $N_c$ quarks, becomes
irrelevant, and fundamental interactions with each quark may be
categorized in terms of the {\it one-body} operators classified by
spin-flavor:
\begin{eqnarray}
J^i & \equiv & \sum_\alpha q^\dagger_\alpha \left( \frac{\sigma^i}{2}
\otimes \openone \right) q_\alpha , \nonumber \\
T^a & \equiv & \sum_\alpha q^\dagger_\alpha \left( \openone \otimes
\frac{\lambda^a}{2} \right) q_\alpha , \nonumber \\
G^{ia} & \equiv & \sum_\alpha q^\dagger_\alpha \left(
\frac{\sigma^i}{2} \otimes \frac{\lambda^a}{2} \right) q_\alpha ,
\label{onebody}
\end{eqnarray}
where the index $\alpha$ sums over the $N_c$ quarks, $\sigma^i$ are
Pauli spin matrices, and $\lambda^a$ are Gell-Mann flavor matrices.
Each distinct operator may be written as a monomial in $J$, $T$, and
$G$ of total order $n$ (with $0 \! \le n \le N_c$) and is termed an
{\it n-body operator}.  A large subset of operators constructed in
this way are redundant or give vanishing matrix elements due to
group-theoretical constraints; the {\it operator reduction rules\/}
derived in Ref.~\cite{DJM} show how to remove systematically all such
operators acting upon the ground-state band.  Since each interaction
requires a factor of $\alpha_s \! = \!  O(1/N_c)$, operators composed
of multiple one-body operators tend to be suppressed in powers of
$1/N_c$.  However, for the low-lying states in the ground-state band
($N$, $\Delta$, etc.), $I^a \, (\equiv T^a \ {\rm for } \ a = 1,2,3)$,
$J^i$, and $G^{i8}$ have matrix elements of $O(N_c^0)$ while $G^{ia}$
with $a = 1,2,3$ and $a = 4,5,6,7$ give matrix elements of $O(N_c^1)$
and $O(N_c^{1/2})$, respectively, $T^a$ with $a = 4,5,6,7$ gives
$O(N_c^{1/2})$, and $T^8$ gives a part $O(N_c^1)$ times the identity
operator (hence redundant) plus a part at $O(N_c^0)$ proportional to
strangeness.

The original KSM theorem~\cite{KapSavMan} (which was originally
applied to nucleon-nucleon scattering) shows that amplitudes with
$|I_t \! - \!  J_t| \! = \! n$ scale at most as $O(1/N_c^n)$.  The
original KSM proof writes $t$-channel exchanges in terms of the
one-body operators, and uses the fact that the only 2-flavor operator
with $O(N_c^1)$ matrix elements is $G^{ia}$.  If the indices on a
string of $G$'s are summed, the operator reduction rules always turn
out to generate a composite operator with subleading $N_c$ counting;
therefore, the leading operators are ones for which the spin and
isospin indices on the $G$'s (of which there are equal numbers) are
unsummed and symmetrized.  A collection of ${\cal J}$ $G$'s combined
in this way thus gives a tensor with $I_t \! = \!  J_t \! = \! {\cal
J}$.  Each contraction or one-body operator $I^a$ or $J^i$ instead of
a $G^{ia}$ costs a relative factor $N_c$, and therefore operators with
$|I_t \! - \! J_t| \! = \!  n$ are suppressed by at least a relative
factor of $1/N_c^n$.

This proof was generalized~\cite{ItJt} to three flavors by using (as
noted above) that the isosinglet strangeness-conserving components
$G^{i8}$ and $T^8$ are effectively $O(1/N_c)$ compared to $G^{ia}$
with $a = 1,2,3$ and hence do not spoil the theorem, while the
strangeness-changing operators $T^a, \, G^{ia}$ with $a = 4,5,6,7$
provide a minimum $O(1/N_c^{1/2})$ suppression for each unit of
strangeness change of the baryon, which is just $Y_t$.  $1/N_c$
corrections to both the $I_t \! = \! J_t$ and $Y_t \! = \! 0$ rules
are therefore straightforward to describe, using the operator
formalism.

\section{$1/N_c$ Corrections in the ${\xbf s}$ Channel} \label{Main}

Section~\ref{1Ncorr} shows how to incorporate $1/N_c$ corrections to
meson-baryon scattering amplitudes, via $t$-channel exchanges with
$I_t \! \neq \! J_t$ or $Y_t \! \neq \! 0$. Section~\ref{crosssec}
shows how to cross amplitudes written in terms of $t$-channel
quantities into ones written in terms of $s$-channel quantities.
Apart from managing the exceptionally cumbersome notation, nothing
remains but to merge the two ideas.  The $t$-channel amplitudes $S_{L
L^\prime J_{\phi^{\vphantom\prime}} J_{\phi^\prime} J_t R_t
\gamma^{\vphantom\prime}_t \gamma^\prime_t I_t Y_t}$ (or $S_{L
L^\prime J^{\vphantom\prime}_\phi J_{\phi^\prime} I_t J_t}$ for the
2-flavor case) are suppressed by $N_c^{-|I_t \! - \!  J_t|}$ (for non
strangeness-exchanging processes) or $N_c^{-Y_t/2}$ (for
strangeness-exchanging processes) compared to the leading-order $I_t
\!  = \!  J_t$, $Y_t \! = \! 0$ amplitudes.  Each $t$-channel
amplitude may be inserted directly into Eq.~(\ref{crossreln3}) plus
Eqs.~(\ref{Sbars}) and (\ref{Sbart}) for the 3-flavor case [or just
Eq.~(\ref{crossreln1}) for the 2-flavor case] to give the
corresponding $s$-channel suppressed amplitudes.

This is not to say that one cannot consider $1/N_c$ suppressions of
orders higher than $1/N_c^{|I_t \! - \! J_t|_{\rm max}}$ or
$1/N_c^{Y_t/2}$ for a given process.  Each amplitude $S$ carries a
leading suppression of $1/N_c^\Delta$ with some definite $\Delta$ as
determined by the $I_t \! = \! J_t$ and $Y_t \! = \! 0$ rules;
however, each one may also have subleading contributions
$O(1/N_c^{\Delta + 1})$ that are not discerned by this simpleminded
analysis.  These results may be summarized as just
\begin{eqnarray}
S_{L L^\prime J_{\phi^{\vphantom\prime}} J_{\phi^\prime}
J_t R_t \gamma^{\vphantom\prime}_t \gamma^\prime_t I_t Y_t} & = &
O \left( 1/N_c^{|I_t \! - \! J_t|} \right) \ \ (Y_t \! = \! 0) \, ,
\nonumber \\ & = &
O \left( 1/N_c^{Y_t/2} \right) \ \ (Y_t \! \neq \! 0) \, ,
\nonumber \\ & \to &
\sum_{{\rm all \; except \,} L, L^\prime}
S_{L L^\prime S S^\prime J_s R_s
\gamma^{\vphantom\prime}_s \gamma^\prime_s I_s Y_s} \ ,
\end{eqnarray}
with quantum numbers $S, S^\prime, J_s, R_s,
\gamma^{\vphantom\prime}_s, \gamma^\prime_s, I_s, Y_s$ for the
amplitudes on the right-hand side limited to those allowed by the
group-theoretical constraints of Eqs.~(\ref{Sbars}) and
(\ref{crossreln3}).

In the case of scattering with spinless pions
($S_{\phi^{\vphantom\prime}} \! = \! S_{\phi^\prime} \! = \! 0$,
$I_{\phi^{\vphantom\prime}} \! = \! I_{\phi^\prime} \! =
\! 1$), Eq.~(\ref{crossreln2}) reduces to the forms used to study
the phenomenology of $\pi N \! \to \! \pi N, \, \pi \Delta$ scattering
processes in Refs.~\cite{CDLN2,CLchiral}.  The amplitude $S_{LL^\prime
S_{B^{\vphantom\prime}} S_{B^\prime} I_s J_s}$ receives a correction
\begin{eqnarray}
-\frac{1}{N_c^{|I_t \! - \! J_t|}} s_{I_t L L^\prime}^{t (J_t \! - \!
I_t)} & = &
(-1)^{L + L^\prime} (9 [L][L^\prime]
[S_{B^{\vphantom\prime}}]^2 [S_{B^\prime}]^2)^{-1/4} [I_t][J_t]
S_{LL^\prime LL^\prime I_t J_t} \ .
\end{eqnarray}

Finally, we present one explicit example of the formalism that has not
previously been considered in the literature: $1/N_c$ corrections to
the process $\pi N \! \to \! \rho N$.  As seen in Ref.~\cite{KL1}, the
processes $\pi N \! \to \! \pi \pi N$ are dominated for large $N_c$ by
resonant $\pi N \! \to \! \pi \Delta$, $\rho N$, or $\omega N$
intermediate states, and moreover, branching fractions for such
processes have been extracted from raw scattering data.  However,
Ref.~\cite{KL1} worked only with the leading $[O(N_c^0)]$ amplitudes
and found the results in many cases (for predicting branching ratios
of given baryon resonances to particular final states) to be rather
inconclusive.  In addition to the large uncertainties in the data, a
principal culprit for this imprecision lay in the omission of $1/N_c$
corrections.  The $1/N_c$-suppressed amplitudes are clearly
significant, because they were shown in several cases to be of the
right order of magnitude to explain discrepancies between data and the
leading-order predictions.  A full reanalysis of the sort performed in
Ref.~\cite{KL1} but including $1/N_c$ corrections is of course far
outside of our current scope, but we can at least show how quickly the
onerous expressions obtained above simplify for a physical case.

Consider two simple cases, both with $I_s \! = \! J_s \! = \! \frac 1
2$ and the initial $\pi N$ in a state of relative $L \! = \! 0$.  Then
the final $\rho N$ can either be in a state of relative $L^\prime \! =
\! 0$ when $S^\prime \! = \! \frac 1 2$ (the $S_{11}$ partial wave),
or $L^\prime \! = \! 2$ when $S^\prime \! = \! \frac 3 2$ (the
$SD_{11}$ partial wave).  Then Eq.~(\ref{crossreln2}) gives
\begin{eqnarray}
S_{11}^{(\pi N)(\rho N)_1} & = & +\sqrt{\frac 3 2} S_{000111} +
\frac{1}{2} S_{000101} \, , \nonumber \\
SD_{11}^{(\pi N)(\rho N)_3} & = & -\sqrt{\frac 3 2} S_{020111} -
\frac{1}{2} S_{020101} \, ,
\end{eqnarray}
where we use the notation of Ref.~\cite{KL1}: The superscript is
$2S^\prime$.  As a reminder, the last two indices of amplitudes $S$ on
the right-hand side are $I_t \, J_t$, so the second amplitude in each
case is $1/N_c$ suppressed.  More amplitudes arise for higher spins
and higher partial waves, but like this example, the explicit
expressions tend to be quite simple in general.

As discussed above, if one is not concerned with the specific SU(3)
quantum numbers $R_s \gamma_s$, one may apply Eq.~(\ref{crossreln3})
directly.  As an example, consider $KN$ scattering; for spinless
mesons, $S \! = \! S_B$ and $S^\prime \! = \!  S_B^\prime$, and the
$s$-channel amplitude $\bar S_{LL^\prime S S^\prime I_s J_s}$ is
denoted in the literature by $(LL^\prime)_{I_s,2J_s}$.  When
specialized to the $S$-wave case ($L \! = L^\prime \! = \! 0$),
Eq.~(\ref{crossreln3}) simply gives
\begin{eqnarray}
\bar S_{000000} & = & \frac{1}{\sqrt{2}} \left( S_{01} + 3 S_{11}
\right) \, , \label{KN1st} \\
\bar S_{000010} & = & \frac{1}{\sqrt{2}} \left( S_{01} - S_{11}
\right) \, . \label{KN2nd}
\end{eqnarray}
Recalling that the last two indices of the $t$-channel amplitudes on
the left-hand side are $I_t$ and $J_t$, we note that Eq.~(\ref{KN1st})
is $O(N_c^0)$ and Eq.~(\ref{KN2nd}) is $O(1/N_c)$.  From the second of
these it follows that $S_{01} \! = \! S_{11}$ up to $O(1/N_c)$
corrections.  In fact, available partial-wave data~\cite{SAID}
supports this approximate equality: Both the real and imaginary parts
of $S_{01}$ and $S_{11}$ have the same signs and basic shapes as
functions of $s$, and are approximately equal for large $s$, with
differences in the 1.5--2.0~GeV region that can be attributed to
relative $O(1/N_c)$ corrections.

\section{Conclusions} \label{concl}

In this paper we have examined in detail the procedure for crossing
between $s$- and $t$-channel quantum numbers for meson-baryon
scattering in the context of the $1/N_c$ expansion.  The $s$-channel
quantum numbers are the ones used most frequently for describing
physical scattering processes.  However, the $t$-channel quantum
numbers are the ones most convenient for quantifying $1/N_c$ power
suppressions, using the degree of violation of the $I_t \! = \! J_t$
or $Y_t \! = \! 0$ rules.

We have given explicit expressions for crossing any given amplitude in
the $t$ channel in terms of a linear combination of amplitudes in the
$s$ channel and vice-versa, in the case of two quark flavors.  In the
3-flavor case, unless one assumes SU(3) symmetry for all meson-baryon
amplitudes, one obtains not a single closed-form crossing solution,
but a series of linear equations that impose constraints on the
amplitudes.

Finally, we have shown how the complicated expressions obtained here
simplify to those used previously, and exhibited as an explicit
example a simple novel case, $\pi N \! \to \! \rho N$ formed in the
$S_{11}$ channel.  Such expressions as obtained here may be used in a
detailed phenomenological analysis, including subleading $1/N_c$
corrections, for processes such as $\pi N \! \to$ multi-$\pi N$ or $K
N$ scattering.

\section*{Acknowledgments}
We thank Tom Cohen for valuable discussions.  This work was supported
by the NSF under Grant No.\ PHY-0456520.

\end{document}